\documentclass[aps,prl,twocolumn,groupedaddress,showpacs,%
amsmath,preprintnumbers]{revtex4}
\input{amssym}
\usepackage{hyperref} 
\def\p {\partial}
\def\t {\tilde}
\def\be {\begin{equation}}
\def\ee  {\end{equation}}
\def\bea {\begin{eqnarray}}
\def\eea {\end{eqnarray}}
\def\nn {\nonumber}
\begin{document}
\preprint{gr-qc/04}
\title{Quantum resolution of black hole singularities}
\author{Viqar Husain$^\dagger$ and Oliver Winkler$^\dagger$}
\email[]{husain@math.unb.ca, oliver@math.unb.ca}
\affiliation{$^\dagger$ Department of Mathematics and Statistics,
University of New Brunswick, Fredericton, NB, Canada E3B 5A3\\
$^*$Perimeter Institute for Theoretical Physics, Waterloo, ON, Canada N2L 2W9}
\date{\today}
\begin{abstract}

We study the classical and quantum theory of spherically symmetric spacetimes with 
scalar field coupling in general relativity. We utilise the canonical formalism 
of geometrodynamics adapted to the Painleve-Gullstrand coordinates, and present 
a new quantisation of the resulting field theory. We give an explicit 
construction of operators that capture curvature properties of the spacetime 
and use these to show that the black hole curvature singularity is avoided in the 
quantum theory.  
  
\end{abstract}

\maketitle

It is well known that many solutions of classical general relativity have curvature 
singularities. The most commonly encountered are the initial or final singularities 
in cosmological models, and the singularity inside a black hole. Just as the classical 
singularity of the Coulomb potential is "resolved" by quantum theory, it is 
believed that any candidate theory of quantum gravity must supply a mechanism for 
curvature singularity resolution. 

This question has a long history with the first studies appearing in the late sixties 
using the canonical formulation of general relativity due to Arnowitt, Deser and Misner 
(known as the ADM formalism). The main approach has been (and continues to be) to first 
focus on specific models derived from general relativity by requiring that spacetimes
have some prescribed symmetries. These reductions fall into two general classes, namely 
models that are particle mechanics systems, and those that are still field theories. 
The reduced model is then quantized in the standard Schr\"{o}dinger approach, and the singularity 
resolution question is studied by looking at operators that capture curvature information. 
If the spectra of such operators are bounded, it is taken as an indication of singularity 
avoidance. 
  
The earliest works on singularity avoidance concerned cosmological models. In particular 
the Gowdy cosmologies received attention \cite{mis,bb,vh1}. Surprisingly however, the black hole 
singularity problem has not been studied to date in ADM geometrodynamics, inspite of relatively 
early formulations of the classical canonical theory in spherical symmetry \cite{BCMN,unruh}, 
and more recent work \cite{tk,kuchar}. Furthermore, the spherically symmetric gravity-scalar 
field system is of central importance for a full study of Hawking radiation. A complete quantization 
of this system could give a solution to all the questions and puzzles that arise from the 
semiclassical Hawking radiation calculation, of which singularity resolution is but one. 

Our focus in this paper is the asymptotically flat, spherically symmetric gravity-scalar field system. 
We first formulate the classical canonical field theory adapted to the Painleve-Gullstrand coordinates. 
This formulation appears the most relevant to the gravity-scalar field quantization problem, 
since there is no "extended" Schwarzschild solution with various sectors to worry about, and 
the coordinates are not singular at the horizon. This classical part of our work complements 
the Schwarzschild parametrization given in \cite{tk,kuchar}. 

For quantization, we use a non-standard approach based on recent work by the authors on 
singularity resolution in cosmology \cite{hw1}, which in turn is inspired by the loop space 
representation \cite{loopr}, its application to cosmology \cite{mb}, and by work of 
Halvorson\cite{halv}. In this quantization the configuration and translation operators are represented 
on a Hilbert space, but the momentum operator (the infinitesimal generator of translations) is not 
defined. We construct a class of operators corresponding to phase space functions that are classically 
singular, and show that these operators are bounded.     

We begin by presenting an outline of the classical Hamiltonian theory of the 
spherically symmetric gravity-scalar field model. Since we would like to adapt this system to the 
Painleve-Gullstrand (PG) coordinates, let us recall that the black hole in these coordinates is 
given by the metric
\be 
ds^2 = -dt^2 + \left(dr + \sqrt{\frac{2M}{r}} dt\right)^2 + r^2 d\Omega^2.
\label{PG}
\ee
The spatial metric $e_{ab}$ given by the constant $t$ slices is flat and the extrinsic curvature 
of the slices is 
\be 
K_{ab} = -\sqrt{\frac{2M}{r^3}}\left( e_{ab} - {\frac{3}{2}} n_a n_b\right), 
\label{K}
\ee
where $n^a=x^a/r$ for Cartesian coordinates $x^a$. The black hole mass information is contained 
only in the extrinsic curvature, which in the canonical ADM variables $(q_{ab},\tilde{\pi}^{ab})$ 
determines the momenta $\tilde{\pi}^{ab}$ conjugate to the spatial metric $q_{ab}$. 

The phase space of the model is defined by prescribing a form of 
the gravitational phase space variables $q_{ab}$ and $\tilde{\pi}^{ab}$, together with falloff 
conditions in $r$ for these variables, and for the lapse and shift functions $N$ and $N^a$, such 
that the ADM 3+1 action for general relativity minimally coupled to a massless scalar field  
\be
S = \frac{1}{8\pi G}\int d^3x dt\left[ \tilde{\pi}^{ab}\dot{q}_{ab} + \t{P}_\phi\dot{\phi} 
- N{\cal H} - N^a {\cal C}_a\right]
\label{act}
\ee
is well defined. The constraints arising from varying the lapse and shift are  
\bea
{\cal H} &=& \frac{1}{\sqrt{q}}\left(\tilde{\pi}^{ab}\tilde{\pi}_{ab} -\frac{1}{2} \tilde{\pi}^2 \right) 
                       \sqrt{q}R(q) \nn\\
          & &- 4\pi G (\frac{1}{\sqrt{q}}\tilde{P}_\phi^2 
                       + \sqrt{q}q^{ab}\p_a\phi\p_b\phi) = 0  \\ 
{\cal C}_a &=& D_c\t{\pi}^c_a - \t{P}_\phi\p_a\phi =0,  
\eea
where $\t{\pi}=\t{\pi}^{ab}q_{ab}$. The corresponding conditions for the matter fields 
$\phi$ and $\t{P}_\phi$ are determined by the constraint equations. The falloff conditions imposed 
on the phase space variables are motivated by the PG metric (\ref{PG}), which itself is to be a 
solution in the prescribed class of spacetimes. These  conditions give the following falloff for the gravitational phase space variables 
(for $\epsilon>0$)
\bea
q_{ab} &=& e_{ab} + \frac{f_{ab}(\theta,\phi)}{r^{3/2+\epsilon}} + {\cal O}(r^{-2})  \\
\pi^{ab} &=& \frac{g^{ab}(\theta,\phi)}{r^{3/2}} + \frac{h^{ab}(\theta,\phi)}{r^{3/2+\epsilon}} 
+ {\cal O}(r^{-2}),
\label{pgfo}
\eea  
where $f^{ab},g^{ab},h^{ab}$ are symmetric tensors, $\pi^{ab}=\t\pi^{ab}/\sqrt{q}$, and $q={\rm det}q_{ab}$. 

In this general setting we introduce now the following parametrization for the 3-metric and conjugate 
momentum for a reduction to spherical symmetry:
\bea
q_{ab} &=& \Lambda(r,t)^2\ n_a n_b + \frac{R(r,t)^2}{r^2}\ ( e_{ab} - n_a n_b)\\
\t{\pi}^{ab} &=& \frac{P_\Lambda(r,t)}{2\Lambda(r,t)}\ n^an^b + \frac{r^2 P_R(r,t)}{4R(r,t)}\ (e^{ab} - n^a n^b).
\label{reduc}
\eea
Substituting these into the 3+1 ADM action (\ref{act}) shows that the pairs $(R,P_R)$ and $(\Lambda,P_\Lambda)$ 
are canonically conjugate variables. We note for example the Poisson bracket 
\be 
 \left\{ \int_0^\infty Rf\ dr, e^{i\lambda P_R(r)} \right\} = i2G\lambda f(r) e^{i\lambda P_R(r)},  
\label{bpb}
\ee
which (below) will be the bracket represented in the quantum theory. 
The falloff conditions induced on these variables from (\ref{pgfo}) are the following
\bea
R &=& r +{\cal O}(r^{-1/2-\epsilon}),\ \ \ P_R = r^{-1/2} + {\cal O}(r^{-1-\epsilon})\nn\\
\Lambda &=& 1 + {\cal O}(r^{-3/2-\epsilon}), \ \ \ P_\Lambda = r^{1/2} + {\cal O}(r^{-\epsilon})\nn\\
\phi &=& r^{-1/2} + {\cal O}(r^{-3/2-\epsilon}), \ \ \ P_\phi = r^{1/2} + {\cal O}(r^{-\epsilon}). \nn
\eea
These together with the falloff conditions on the lapse and shift functions
\be 
N^r = r^{-1/2} + {\cal O}(r^{-1/2-\epsilon})\ \ \ N = 1 + {\cal O}(r^{-\epsilon})
\ee
ensure that the reduced ADM 1+1 action  
\bea 
S_R &=& \frac{1}{2G}\int dtdr \left(P_R\dot{R} + P_\Lambda\dot{\Lambda} + P_\phi\dot{\phi}  
    - \cdots \right)\nn\\
 &&+ \int_\infty dt N^r\Lambda P_\Lambda
\eea
obtained by substituting (\ref{reduc}) into (\ref{act}) is well defined. This completes the definition 
of the classical theory. The two reduced constraints, the Hamiltonian and radial diffeomorphism 
generators are first class. The surface term is needed to ensure functional differentiability 
of $S_R$.

At this stage we perform a time gauge fixing using the condition $\Lambda =1$ motivated by PG 
coordinates. This is second class with the Hamiltonian constraint, which therefore must be imposed 
strongly and solved for the conjugate momentum $P_\Lambda$. This gauge fixing therefore eliminates 
the dynamical pair $(\Lambda,P_\Lambda)$, and leads to a system describing the dynamics of 
the variables $(R,P_R)$ and $(\phi,P_\phi)$ \cite{hw2}. The reduced radial diffeomorphism generator 
\be 
P_\Lambda'(R,\phi,P_R,P_\phi) + P_\phi\phi' +P_RR'=0
\label{reddiff}
\ee 
remains as the only first class constraint, which also generates the local dynamics. 

To address the singularity avoidance issue, we first extend the manifold on which the fields 
$R$ etc. live to include the point $r=0$ (ie.  from ${\Bbb R}^+$ to ${\Bbb R}^+_0$), which in the gauge 
fixed theory is the classical singularity. We then ask what classical phase space observables capture curvature information. For homogeneous cosmological models, a natural choice 
is the inverse scale factor $a(t)$. This observable can be represented as an operator in both the 
connection-triad variable based loop representation \cite{mb}, as well as in an ADM variable 
quantization \cite{hw1}. For our model, a guide is provided by the gauge fixed theory without matter 
where it is evident that it is the extrinsic curvature (\ref{K}) that diverges at $r=0$, which is the 
Schwarzschild singularity. This suggests, in analogy with the inverse scale factor, that 
we consider the field variable $1/R$ as a measure of curvature. A more natural choice would be 
a scalar constructed from the phase space variables by contraction of tensors. A simple possibility is 
\be 
\t{\pi} = \frac{1}{2}\left(\frac{P_\Lambda}{R^2} + \frac{P_R}{\Lambda R} \right).
\ee  
The small $r$ behaviour of the phase space variables ensures that any divergence in $\t{\pi}$ is 
due to the $1/R$ factor. We therefore focus on this. A first observation is that the configuration 
variables $R(r,t)$ and $\phi(r,t)$ defined at a single point do not have well defined operator 
realizations. Therefore we are forced to consider phase space functions integrated over 
(at least a part of) space. A functional such as 
\be 
R_f=\int_0^\infty dr fR
\ee 
for a test function $f$ provides a measure of sphere size in our parametrization of the metric. 
We are interested in the reciprocal of this for a measure of curvature. 
Since $R\sim r$ asymptotically, the functions $f$ must have the falloff $f(r) \sim r^{-2-\epsilon}$ 
for $R_f$ to be well defined. Using this, it is straightforward to see that $1/R_f$ diverges 
classically for small spheres: we can choose $f>0$ of the form $f\sim 1$ for $r << 1$, which for large 
$r$ falls asymptotically to zero. Then $R_f\sim r^2$ and $1/R_f$ diverges classically for small spheres. 

A question for the quantum theory is whether $1/R_f$ can be represented densely on a Hilbert space 
as $1/\hat{R}_f$.  This is possible only if the chosen representation is such that $\hat{R}_f$ does 
not have a zero eigenvalue. If it does, we must represent $1/R_f$ as an operator more indirectly, 
using another classically equivalent function. Examples of such functions are provided by Poisson 
bracket identities such as 
\be 
\frac{1}{|R_f|} = \left(\frac{2}{iG f(r)}\ e^{-iP_R(r)}\left\{\sqrt{|R_f|}, e^{iP_R(r)} \right\}\right)^2,
\label{invRf}
\ee
as the functions $f$ do not have zeroes. The representation for the quantum theory described below 
is such that the operator corresponding to $R_f$ has a zero eigenvalue. Therefore we represent $1/R_f$ 
using the r.h.s. of (\ref{invRf}).   The central question for singularity resolution is whether the 
corresponding operator is densely defined and bounded \cite{fullqg}. 

We now proceed with a quantization of the 1+1 dimensional scalar theory 
of the fields $R$ and $\phi$. A guideline is the idea to turn an expression 
like $e^{-ip}\{x^{\frac{1}{2}},e^{ip}\}$, which resembles Eqn. (\ref{invRf}), 
into a quantum operator that has eigenvectors that correspond to the classical singularity. 
It turns out that the implementation of this idea requires that plane waves be normalizable 
states in the quantum theory. This renders a standard Schr\"{o}dinger type quantization unsuitable. 
Instead one is forced to use a quantization based on an enlarged quantum configuration space. 
This type of quantization has already been fruitfully applied to an investigation of the initial
big bang singularity in the FRW cosmological model \cite{hw1}, where there is  
a discussion of some technical aspects of the quantization. We describe here the 
ingredients necessary for our discussion of the black hole singularity. 

As a prelude, let us consider the one-dimensional non-relativistic particle with 
Hamiltonian $H(x,p)$. The basis states are  $|x\rangle\equiv |{\rm exp} (ix p)\rangle$ with 
inner product $\langle x | y\rangle = \delta_{x,y}$, with the Kronecker delta indicating
normalizability (as opposed to the case of a delta function in standard Schr\"{o}dinger quantum
mechanics). The enlarged quantum configuration space is the so-called Bohr compactification of
$\Bbb R$ (named for mathematician Harald Bohr; see \cite{halv,hw1} for details). Position operators act 
multiplicatively, $\hat{x} |x\rangle = x |x\rangle$, while {\it momentum operators exist only in 
their exponentiated form} as finite  shift operators $\widehat{e^{iy p}} |x\rangle = |x-y\rangle$. 
It should be stressed that the apparent discreteness evident here is not the same as the one due 
to introduction of a fixed space lattice at the classical level. 
 
The quantization of free scalar fields in this framework is similar, and has been developed 
in \cite{TT98,ALS}. A quantum field is characterized by its excitations at $N$ points in space. 
The important difference from standard quantum field theory is that here such states are 
normalisable.  A typical basis state is  
\be 
|e^{i \sum_k a_k P_R(x_k)}, e^{i L^2\sum_l b_l P_\phi(y_l)}\rangle
\equiv |a_1\ldots a_{N_1};b_1\ldots b_{N_2}\rangle, 
\label{basis}
\ee
where the factors of $L$ in the exponents reflect the length dimensions of the respective field 
variables, and $a_k,b_l$ are real numbers which represent the excitations of the scalar quantum fields 
$R$ and $\phi$ at the radial locations $\{x_k\}$ and $\{y_l\}$. The inner product on this 
basis is  
\bea
&&\langle a_1 \ldots a_{N_1};b_1,\ldots b_{N_2}|a'_1 \ldots a'_{N_1}; b'_1\ldots b'_{N_2} \rangle \nn\\
&& = \delta_{a_1,a_1'}\ldots \delta_{b_{N_2},b_{N_2}'},\nn 
\eea
if the states contain the same number of sampled points, and is zero otherwise. 

The action of the basic operators are given by
\bea 
&&\hat{R}_f\ |a_1 \ldots a_{N_1};b_1 \ldots b_{N_2}\rangle = \nn\\
&&L^2 \sum_k a_k f(x_k)|a_1 \ldots a_{N_1};b_1 \ldots b_{N_2}\rangle,\\
&& \widehat{e^{i \lambda_j P_R(x_j)}}|a_1\ldots a_{N_1};b_1\ldots b_{N_2}\rangle
 = \nn\\
&&|a_1\ldots, a_j-\lambda_j,\ldots a_{N_1};b_1\ldots b_{N_2}\rangle,
\eea
where $a_j$ is $0$ if the point $x_j$ is not part of the original basis state. In this case the
action creates a new excitation at the point $x_j$ with value
$-\lambda_j$. These definitions give the commutator  
\be 
\left[\hat{R}_f,\widehat{e^{i\lambda P_R(x)}} \right] = -\lambda f(x) L^2 \widehat{e^{i\lambda P_R(x)}}.
\ee
Comparing this with  (\ref{bpb}), and using the Poisson bracket commutator correspondence,  
we see that $L = \sqrt{2} l_P$, where $l_P$ is the Planck length. There are similar operator 
definitions for the canonical pair $(\phi,P_\phi)$. 

Using the expressions for the basic field operators, we can construct an operator
corresponding to a classical singularity indicator:
\bea
&&\widehat{\frac{1}{|R_f|}}\    |a_1\ldots a_{N_1};b_1\ldots b_{N_2}\rangle \nn\\
&=&  \left( \frac{2}{l_P^2 f(x_j)}\widehat{e^{- iP_R(x_j)}}\ \left[ \widehat{\sqrt{|R_f|}},
\ \widehat{e^{iP_R(x_j)}} \right] 
\right)^2 \nn\\
&&\times |a_1 \ldots a_{N_1};b_1\ldots b_{N_2}\rangle.
\eea
One can show that the basis states are eigenvectors of this operator, and that all resulting 
eigenvalues are bounded. We illustrate this with the state  
\be
|{\rm S}_{a_0}\rangle \equiv|e^{ia_0P_R(r=0)}\rangle,
\ee
which represents an excitation $a_0$ of the quantum field $\hat{R}_f$ at 
the point of the classical singularity:
\bea
\hat{R}_f |{\rm S}_{a_0}\rangle &=& (2 l_P^2) f(0) a_0\ |{\rm S}_{a_0}\rangle, \\
\widehat{\frac{1}{|R_f|}}|{\rm S}_{a_0}\rangle  &=& 
\frac{2}{l_P^2 f(0)}\left(|a_0|^{1/2} - |a_0-1|^{1/2}\right)^2 |{\rm S}_{a_0}\rangle  \nn
\eea
which is clearly bounded. This shows that the singularity is resolved at the quantum level.  
In particular if there is no excitation of $R_f$ at the classical singularity, ie. $a_0=0$, 
the upper bound on the eigenvalue of the inverse operator is $2/l_P^2$. This fact also 
gives an interesting result concerning areas of the symmetry 2-spheres: {\it The inverse area
has an upperbound proportional to $1/l_p^2$}. 

Having established that the curvature spectrum is bounded at the kinematical level, 
let us ask if this conclusion is affected by dynamics. This involves considering the 
action of the remaining constraint (\ref{reddiff}) on basis states (\ref{basis}). 
Our strategy is to show that this action is well defined even on states of zero eigenvalue 
of $\hat{R}_f$, which is the indication that quantum evolution continues without divergence  
to the location of the classical singularity. This property is also the content of "dynamical 
singularity resolution" in earlier works on cosmology \cite{hw1,mb}. It shows that local 
quantum evolution is well defined, unlike its classical counterpart.   

The reduced constraint (\ref{reddiff}) and the evolution equations 
derived from it \cite{hw2} contain factors of $1/R$, which are the only 
sources of classical divergence at $R=0$. This constraint is quantised by representing these 
factors by powers of the $\widehat{1/R_f}$ operator defined above. The other terms in the 
constraint contain the field variables and their conjugate momenta. The former are diagonal,  
and the latter are represented as shift operators \cite{hw3}. In both cases their action on a basis 
state is finite. Thus the only source of classical divergence in the constraint gets  
resolved. This may be compared to the resolution of the Coulomb divergence in atomic physics 
where the expectation value of $V(r)\sim 1/r$ is finite and bounded above, although the details of 
the mechanism are different. There it provides a lower bound on energy. In our case it provides 
dynamical curvature singularity resolution via finite action of the constraint operator on 
states with zero eigenvalue of $\hat{R}_f$.  

The second aspect of dynamical singularity resolution concerns the surface term, which may be 
viewed as generating dynamics at spatial infinity. This "true" Hamiltonian is given by 
$N^rP_\Lambda(R,\phi,P_R,P_\phi)$ evaluated at the point $r=\infty$ (since the sphere integrals 
just give a $4\pi$ factor). As before, this contains a factor $1/R$ which is bounded above at the
quantum level for the same reasons given above. Therefore $\langle \psi| \hat{P_\Lambda} |\psi\rangle$ 
is bounded for basis states $|\psi\rangle$, and so is the  expectation value in time evolved states 
$\langle \psi| \hat{P_\Lambda} \widehat{1/R_f} \hat{P_\Lambda} |\psi\rangle$. 
    
Let us summarise these results and compare them with what has been achieved for cosmology. 
We  presented a quantization of the gravity-scalar field system in which 
there are operators corresponding to classical curvature. All such operators contain a 
$\widehat{1/{R}_f}$ factor, which we have shown to be bounded. We have shown that the upper 
bound on curvature is unaffected by time evolution via the remaining constraint, 
which encapsulates the quantum dynamics of the system. This establishes both kinematical and 
dynamical resolution of black hole singularities.  

The work on singularity resolution in homogeneous and isotropic cosmology in both the 
connection \cite{mb} and metric variable \cite{hw1} approaches does not utilise a time 
gauge fixing. The action of the Hamiltonian constraint is studied on a kinematical 
Hilbert space, and dynamical singularity resolution is established by noting  
that the quantum constraint equation allows evolution past the state corresponding to 
the classical singularity. What we have shown here is that the same holds true in a field 
theory setting with radial inhomogeneity, in a fixed time gauge. This requires careful 
definition of the relevant field theory operators, which is much closer to what is 
expected for full quantum gravity. 

Conceptually, our work rests on the use of PG coordinates in a field theory 
setting with matter. The use of PG coordinates permits a unified treatment of the singularity 
resolution problem in a matter coupled setting. In contrast quantising the source free 
black hole in Schwarzschild coordinates requires a separation of the interior "cosmological" and 
"exterior" asymptotically flat regimes \cite{modes}. This advantage will be useful for 
a study of the full quantum collapse problem.  
 
The quantisation also provides a setting for studying other operators such as null geodesic 
expansions \cite{hw3}. These permits a definition of black holes in quantum theory  
without recourse to classical "horizon boundary conditions" at the classical level, which 
has the potential to address Hawking radiation in a fully quantised setting. 
 
\acknowledgments{This work was supported in part by the Natural Science and Engineering 
Research Council of Canada}

\end{document}